\documentclass{aa}  
\usepackage{graphicx}
\usepackage{natbib}
\usepackage{txfonts}
\usepackage{amssymb}
\usepackage{verbatim}
\newcommand{\chandra}{{\em Chandra}}

\begin{document}

   \title{The nearest neighbor statistics for X-ray source counts\\
          I.The method}

   \author{A. M. So\l tan         \inst{}                }

   \offprints{A. M. So\l tan}

   \institute{Nicolaus Copernicus Astronomical Center,
              Bartycka 18, 00-716 Warsaw, Poland\\
              \email{soltan@camk.edu.pl}                 }

   \date{Received ~~ / Accepted }

  \abstract
   {Most of the X-ray background (XRB) is generated by discrete X-ray
sources.  It is likely that still unresolved fraction of the XRB is
composed from a population of the weak sources below the present
detection thresholds and a truly diffuse component. It is a matter of
discussion a nature of these weak sources.}
   {The goal is to explore the effectiveness of the nearest neighbor
statistics (NNST) of the photon distribution for the investigation of
the number counts of the very weak sources.}
   {All the sources generating at least two counts each induce a
nonrandom distribution of counts. This distribution is analyzed by
means of the NNST. Using the basic probability equations, the
relationships between the source number counts $N(S)$ and the NNST are
derived.}
   {It is shown that the method yields constraints on the $N(S)$
relationship below the regular discrete source detection threshold.
The NNST was applied to the medium deep \chandra\ pointing to assess
the source counts $N(S)$ at flux levels attainable only with the very
deep exposures. The results are in good agreement with the direct
source counts based in the \chandra\ Deep Fields (CDF).}
   {In the next paper of this series the NNST will be applied to the
the CDF to assess the source counts below the present flux limits.}


   \maketitle
%

\section{Introduction}

The X-ray background (XRB) is mostly generated by discrete extragalactic
sources (e.g. \citealt{lehmann01, kim07}, and references therein).  Thus, the
source counts provide the essential information on the constituents of the
XRB and the X-ray $N(S)$ relationship has been a subject of numerous studies
for the last $40$ years.

The individual point-like source is detected if a number of counts
within a specified area exceeds the assumed threshold. Size of the
detection box is defined by the Point Spread Function (PSF), while the
detection threshold is usually selected to minimize number of false
detections and at the same time to maximize number of the real sources.
The detection threshold is typically set at the level of $4-5\, \sigma$
above the local average count density.  A presence of weaker sources,
below the formal detection threshold, is manifested by the increased
fluctuations of the count distribution as compared to the fluctuations
expected for the random counts.

A common approach to assess counts of sources weaker than the detection
limit is based on the count density fluctuation analysis. To quantify
signal generated by the discrete sources one should determine the
intensity distribution $P(D)$, i.e. the histogram of the number of
pixels as a function of the number of counts. The observed function
$P(D)$ is then compared with the functions obtained from the simulated
count distributions \citep[e.g.][]{hasinger93,miyaji02}.  It is assumed
that the simulated source counts represent the actual source
distribution if the model $P(D)$ function mimics the observed histogram.
A contribution of point sources to the count distribution one can
estimate also using the auto-correlation function (ACF). Since the
integral of the ACF is directly related to the second moment of the
$P(D)$ distribution, \citep[e.g.][]{soltan91} both methods are closely
related.

An innovative method to assess the number of weak sources was proposed
by \citet{georgakakis08}. In their approach the count distribution
in the detection cell is explicitly expressed as a sum of the source
and background counts. As a result of a rigorous application of the
Poisson statistics, a flux probability distribution is derived as
a function of the total and background counts observed in the detection
cell. This probability distribution combined with the adequately defined
sensitivity map of a given observation is then used to estimate the
the source number counts.

The count fluctuations are proportional to the source intensities.
Thus, the observed fluctuation amplitude is dominated by the sources
just below the detection threshold set for the individual objects,
whereas it is only weakly sensitive to the fainter sources which
produce smaller number of counts. One should note, however, that every
source which produces more than one count generates deviation from the
random count distribution. In the present paper we investigate the
efficiency of the nearest neighbor statistics (NNST) for the  weak
source analysis and we show that the NNST is a powerful tool to
estimate the source counts, $N(S)$, down to very low flux levels. We
apply this technique to one of the \chandra\ AEGIS fields with the
exposure of $465$\,ks. The NNST allows us to obtain the $N(S)$
relationship extending down to $2\times
10^{-17}$\,erg\,cm$^{-2}$s$^{-1}$ and $7\times
10^{-17}$\,erg\,cm$^{-2}$s$^{-1}$in the $0.5-2$\,keV and $2-8$\,keV
energy bands, respectively, i.e. a factor of $5-10$ below the standard
sensitivity threshold corresponding to this exposure. Since the
present count estimates are contained within the flux range covered by
the direct source counts derived from the deepest \chandra\ fields
(such as CDFS), the effectiveness of the NNST method could be directly
assessed.

The organization of the paper is as follows. In the next section, the
method and all the relevant formulae are presented. Then, in
Sect.~\ref{observations}, the observational material is described and
the computational details including questions related to the PSF are
given.  Results of the calculations, i.e.  estimates of the source
counts below the nominal sensitivity limit are presented in Sect.~4. The
results are summarized and discussed in Sect.~5. Prospects for the
application of the NNST to the deep \chandra\ fields are presented.

\section{The nearest neighbor statistics \label{nnst}}

In this section a general formulae based on the theory of probability
are derived, while all the details related to the actual observations
(e.g.  relationship between counts and photon energy, instrument
sensitivity) are discussed in Sect.~\ref{observations}.

In the present consideration we assume that counts in the given
\chandra\ ACIS\footnote{In all the analysis we use ACIS-I chips 0-3.}
observation are distributed according to the following model: some
a priori unknown fraction of counts is randomly distributed, i.e. the
positions of the counts are fully described by the Poisson statistics,
while all the remaining photons are clustered in the point-like sources.
Such model corresponds to the observation obtained using the idealized
telescope without vignetting and detector with perfectly uniform
response over the field of view. The real telescope-detector combination
introduces numerous deformations to this ideal picture. We address this
question below.

We further assume, that the positions of sources are also randomly
distributed.  Smoothly distributed counts constitute physically
heterogeneous collection which contains both the particle background and
various components of the foreground X-ray emission including scattered
solar X-rays, the geocoronal oxygen lines as well as the thermal
emission of hot plasma within our Galaxy
\citep[e.g.][]{hasinger92,galeazzi07,henley07} and the emission by the
WHIM \citep[e.g.][]{soltan07}. Truly diffuse extragalactic background
\citep{soltan03} and weak discrete sources, each producing in the final
image exactly one photon, contribute also to these counts. A separation
of the extragalactic component from all the counts can be achieved in
statistical terms using spectral information, however, individual event
cannot be definitely classified as local or extragalactic.

Discrete sources in the deep X-ray exposures are predominantly
extragalactic. Nevertheless, galactic sources are potentially present in
the data and are included in the calculations.  Photons coming from a
discrete sources are distributed in the image in clumps defined by the
PSF of the telescope.

Within the present model, statistical characteristics of the count
distribution are fully defined: there are two population of events, the
first is randomly distributed and the second is concentrated in the PSF
shaped clusters. This feature of the count distribution is conveniently
formulated using the NNST. Let $n_{\rm t}$ denotes the total number of
counts in the investigated field, $n_1$ - the number of events
distributed randomly, or ``single photons'', $n_2$ - the number of
photons due to sources producing each exactly 2 photons, $n_3$ - number
of photons due to ``three photon'' sources and so on. Thus,

\begin{equation}
\label{counts}
n_1 + n_2 + ... + n_k + ... + n_{k_{\rm max}} = n_{\rm t}\,,
\end{equation}
where the left hand side sum extends over all the sources and $k_{\rm
max}$ is the number photons produced by the brightest source in the
field. The number of photons $n_k$ is related to the number of sources
in an obvious way: $k\,N(k) = n_k$, where $N(k)$ denotes the number of
``$k$-photon sources''.

Using the basic relationships of the probability theory, one can
calculate $P(r)$ -- the probability that the distance to the nearest
neighbor of a randomly chosen event exceeds $r$:

\begin{equation}
\label{basic}
p_1 P(r\!\mid\!1) + p_2 P(r\!\mid\!2) + 
       ... + p_{k_{\rm max}} P(r\!\mid\! k_{\rm max}) = P(r)\,,
\end{equation}
where $p_k$ denotes the probability that the randomly chosen photon is
produced by the ``$k$-photon source'' ($k = 1, 2, ..., k_{\rm max}$),
and $P(r\!\mid\!k)$ is the conditional probability that there are no
other counts within $r$ provided the selected event belongs to the
$k$-photon source.

The probability $P(r)$ can be estimated for the given distribution of
counts by measuring the distance to the nearest neighbor for the each
photon in the field. Similarly, $P(r\!\mid\!1)$ is given by the
distribution of distances to the nearest photon from the randomly
distributed points. Assuming that the distribution of ``single counts''
is not correlated with the distribution of photons from $k\ge 2$
sources, and that sources are distributed randomly, the probability
$P(r\!\mid\!k)$ for $k\ge 2$ are related to $P(r\!\mid\!1)$ and to the
PSF by the expression:

\begin{equation}
\label{psf1}
P(r\!\mid\! k) = P(r\!\mid1) \cdot {\cal P}(r\!\mid\! k)\,,
\end{equation}
where ${\cal P}(r\!\mid\! k)$ is the probability that the distance from
the randomly chosen photon produced by the $k$-photon source to its
nearest neighbor from the same source exceeds $r$. This quantity is
fully defined by the PSF.

To estimate the probabilities $p_k$ we now use a ratio $n_k/n_{\rm t}$
and the Eq.~\ref{basic} takes the form:

\begin{equation}
\label{basic1}
\frac{n_1}{n_{\rm t}}\, P(r\!\mid\!1)\, +\, \sum_{k=2}^{k_{\rm max}}\,
   \frac{n_k}{n_{\rm t}}\, P(r\!\mid\!1)\,{\cal P}(r\!\mid\! k) = P(r)\,.
\end{equation}

It is worth to note that Eq.~\ref{basic1} is linear in the photon counts
$n_k$, what makes it particularly suitable for the estimates of the
contribution of weak sources to the total counts. Both the second moment
of the count distribution in pixels and the autocorrelation function
depend on squares of the photon counts. Substituting successive  values
of $r$ into Eq.~\ref{basic1}, a set of linear equations is constructed
which allows us to estimate the unknown counts $n_k$.

In the deep \chandra\ observation a large number of relatively strong
sources is detected and the range of source fluxes is much too wide to
apply Eq.~\ref{basic1} in the form given above where the source fluxes
are listed consecutively from $k = 2$ up to some maximum value of
$k_{\rm max}$ representing the strongest source in the field. Since we
are interested in the counts at the faint end (down to $k=2$), the value
of $k_{\rm max}$ is selected at the conventional detection limit.  All
the sources above this threshold are pinpointed and removed from the
data and the subsequent analysis is concentrated on the sources which
cannot be individually recognized.

One can express the number of photons due to $k$-photon sources by
means of the differential source counts $N(S)$:

\begin{equation}
\label{counts1}
n_k = k\,\int_{S_{\rm min}}^{S_{\rm max}} {\rm d}S\,N(S)\,
                                        \mathfrak{P}(k\!\mid\!S)\,,
\end{equation}
where $\mathfrak P(k\!\mid\!S)$ is the probability, that the source
generating flux $S$ delivers $k$-photons, while the integration limits
$S_{\rm min}$ and $S_{\rm max}$ define the full range of the source
fluxes.  It is convenient to introduce the instrumental count as a flux
unit. The flux $s$ expressed in the ACIS counts in the definite
observation is related to the flux in physical units, $S$, by:

\begin{equation}
\label{def_cf}
s = S / {\rm cf}\,,
\end{equation}
where ${\rm cf}$ is the conversion factor which has units of
``erg\,cm$^{-2}$\,s$^{-1}$/\,count''
and is related to the parameter ``exposure map'' defined in a standard
processing of the ACIS data\footnote{See http://asc.harvard.edu/ciao.
For the real observations, both the cf and exposure map are functions of
the position. At this stage these parameters are assumed constant.}:
exposure map $ = {\rm cf}\cdot\langle E \rangle$, where $\langle E
\rangle$ denotes the average photon energy.

For the power law source counts, $N(s)= N_o\,s^{-b}$, we have:

\begin{equation}
\label{gamma}
n_k = N_o\:\frac{\Gamma(k-b+1,s_{\rm min}) - \Gamma(k-b+1,s_{\rm max})}
                {\Gamma(k)}\,.
\end{equation}

If the slope of the counts $b$ is constant over a sufficiently wide range of
fluxes (i.e. $s_{\rm min} << 2$ and $s_{\rm max} > k_{\rm max}$), one
might replace the integration limits $s_{\rm min}$ and $s_{\rm max}$ by
$0$ and $\infty$, respectively, to get:

\begin{equation}
\label{gamma1}
n_k = N_o\:\frac{\Gamma(k-b+1)}{\Gamma(k)}\,.
\end{equation}

Thus, for the source counts represented by a single power law over
sufficiently wide range of fluxes, Eq.~\ref{basic1} takes the form:

\begin{equation}
\label{basic2}
\frac{n_1}{n_{\rm t}}\,P(r\!\mid\!1)\:+\:\frac{N_o}{n_{\rm t}}\,
      \sum_{k=2}^{k_{\rm max}} \frac{\Gamma(k-b+1)}{\Gamma(k)}\:
      P(r\!\mid\!1)\,{\cal P}(r\!\mid\!k) = P(r)\,.
\end{equation}

The extension of this expression for the source counts with varying
slope is straightforward. In particular, for the broken power law,
the function $\Gamma(k-b+1)$ is replaced by a proper combination
of the incomplete gamma functions.
Substituting 

\begin{equation}
n_1 = n_{\rm t} - \sum_{k=2}^{k_{\rm max}}\, n_k
\end{equation}
and using Eq.~\ref{gamma1} we finally get:
\begin{equation}
\label{final}
\frac{N_o}{n_{\rm t}}\, P(r\!\mid\!1)
     \sum_{k=2}^{k_{\rm max}} \frac{\Gamma(k-b+1)}{\Gamma(k)}\:
     \left[1-{\cal P}(r\!\mid\!k)\right] = P(r\!\mid\!1) - P(r)\,.
\end{equation}

Equation~\ref{final} contains two parameters which fully describe the
power law source counts in the investigated flux range, viz. the
normalization $N_o$ and the slope $b$. Since the normalization at the flux
$s = k_{\rm max}$ is defined by the actual source counts above this
threshold, only the slope remains unknown.
 
\section{Observational material \label{observations}}

To test efficiency of the present method I have selected a set of $16$
close \chandra\ pointings withing the AEGIS\footnote{All-wavelength
Extended Groth strip International Survey, see
http://aegis.ucolick.org/index.html.}.  The observations span a period of
$6$ months and the data have been processed in a uniform way with the
recent pipeline processing versions. The details of $16$ observations used
in the present paper are given in Table~\ref{obs_log}. All the exposures
have been scrutinized with respect to the background flares and only
``good time intervals'' were used in the subsequent analysis.  The data
have been split into two energy bands: S -- soft ($0.5-2$\,keV) and H --
hard ($2-8$\,keV).

\subsection{The exposure map \label{expmap}}

The observations, listed in Table~\ref{obs_log} were merged to create a
single count distribution and exposure map. A circular area covered by
all the pointings with a relatively uniform exposure, centered at ${\rm
RA} = 14^{\rm h}20^{\rm m}12^{\rm s}$, ${\rm Dec} = 53\degr 00\arcmin$
with radius of $6\farcm0$ has been selected for further processing. The
exposure map of the individual observation resulting from various
instrumental characteristics\footnote{To name the most obvious:
vignetting, gaps between chips, telescope wobbling and all kinds of
chip imperfections.}, has a complex structure. In effect, the exposure
map of the merged observation is even less regular and is devoid of any
clear symmetries. On the other hand, a rough texture of the individual
exposure is reduced and smoothed in a sum of $16$ components. To reduce
further the variations of the exposure map over the investigated area,
a threshold of the minimum exposure has been set separately for both
energy band. Pixels below this threshold have not been used in the
calculations.

A threshold has been defined at $\sim\!75$\,\% of the maximum value of
the exposure map for the both energy bands. In effect, the maximum
deviations of the exposure from the average value do not exceed
$15$\,\% and $18$\,\% in the bands S and H, respectively, and the
corresponding rms of the exposures amount to $5.9$ and $6.2$\,\%. In
Table~\ref{exposures} the conversion factors calculated using the
relevant amplitudes of the exposure maps are given. In the calculations
``from counts to flux'' a power spectrum with a photon index
$\Gamma_{\rm ph} = 1.4$ was assumed \citep{kim07}.

Variations of the conversion factor, ${\rm cf}$, over the investigated
area alter the source fluxes via Eq.~\ref{def_cf} and, consequently,
the source counts $N(S)$. For the power law counts the ${\rm cf}$
uncertainty affects the counts normalization and does not change the
slope. It is shown in the Appendix that -- as long as the ${\rm cf}$
variations remain small -- they modify the probability distributions
$P(r\!\mid\!1)$ and $P(r)$ in such a way that the solution of
Eq.~\ref{final} is not affected.

\begin{table}
\caption{The \chandra\ AEGIS observations used in the paper}
\label{obs_log}
\centering
\begin{tabular}{rcccr}
\hline\hline
\noalign{\smallskip}
Obs.&\multicolumn{2}{c}{
             Observation and processing}&Processing & Exposure \\
 ID &\multicolumn{2}{c}{dates}          &version    & time [s] \\
\hline
\noalign{\smallskip}
9450&  2007-12-11  & 2007-12-13 &  7.6.11.3  &   29100 \\
9451&  2007-12-16  & 2008-01-02 &  7.6.11.4  &   25350 \\
9793&  2007-12-19  & 2007-12-21 &  7.6.11.4  &   44750 \\
9725&  2008-03-31  & 2008-04-02 &  7.6.11.6  &   28050 \\
9842&  2008-04-02  & 2008-04-03 &  7.6.11.6  &   19450 \\
9844&  2008-04-05  & 2008-04-06 &  7.6.11.6  &   34600 \\
9866&  2008-06-03  & 2008-06-05 &  7.6.11.6  &   31450 \\
9726&  2008-06-05  & 2008-06-06 &  7.6.11.6  &   39750 \\
9863&  2008-06-07  & 2008-06-07 &  7.6.11.6  &   24100 \\
9873&  2008-06-11  & 2008-06-12 &  7.6.11.6  &   30850 \\
9722&  2008-06-13  & 2008-06-15 &  7.6.11.6  &   19900 \\
9453&  2008-06-15  & 2008-06-17 &  7.6.11.6  &   22150 \\
9720&  2008-06-17  & 2008-06-18 &  7.6.11.6  &   26000 \\
9723&  2008-06-18  & 2008-06-20 &  7.6.11.6  &   30950 \\
9876&  2008-06-22  & 2008-06-23 &  7.6.11.6  &   25050 \\
9875&  2008-06-23  & 2008-06-25 &  7.6.11.6  &   33150 \\
    &   &  &\multicolumn{2}{r}{Total exposure} 464650\,~ \\
\hline
\end{tabular}
\end{table}

\begin{table}
\caption{Energy bands and conversion factors}
\label{exposures}
\centering
\begin{tabular}{lcllll}
\hline\hline
\noalign{\smallskip}
\multicolumn{2}{c}{Energy band}&
                   \multicolumn{4}{c}{Conversion factors$^a$} \\
\multicolumn{2}{c}{~~~~~~~~~[keV]}
               &  Average  & ~~rms     &  minimum    &   maximum  \\
\hline
\noalign{\smallskip}
S  &$0.5 - 2$  & ~~$1.469$ &  $0.087$  & ~~~$1.279$ & ~~~$1.687$ \\
H  & $2   - 8$ & ~~$5.546$ &  $0.346$  & ~~~$4.862$ & ~~~$6.511$ \\ 
\hline
\noalign{\smallskip}
\multicolumn{6}{p{85mm}}{$^a$ The conversion factor (cf) has units of
$10^{-17}\,({\rm erg}\cdot{\rm cm}^{-2}\cdot {\rm s}^{-1})/{\rm count}$.}
\end{tabular}
\end{table}

\subsection{The Point Spread Function \label{psf_sec}}

The \chandra\ X-ray telescope PSF is a complex function of source position
and energy \citep[e.g.][]{allen04}. However, to compute the nearest
neighbor probability distribution for counts generated by a point-like
source, ${\cal P}(r\!\mid\!k)$, we do not need a full model of the PSF
shape. The aim is to find a convenient analytic PSF approximation
adequately reproducing the NNST over the investigated area for each energy
band. The model should be applicable to the merged AEGIS observations
processed in a standard way. 

\begin{figure}
\resizebox{\hsize}{!}{\includegraphics{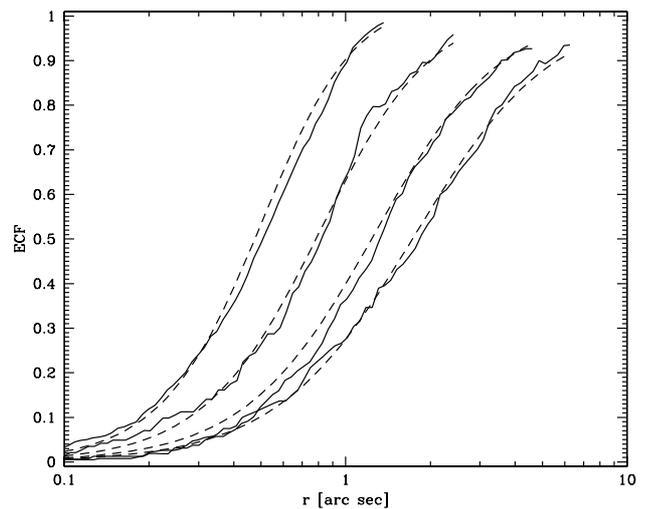}}
\caption{Encircled count fraction (ECF) as a function of distance from
the count centroid. Example distributions are shown for $4$ sources at
$1\farcm 0$, $3\farcm 2$, $5\farcm 0$,
and $6\farcm 1$ from the field center. Solid curves --  observed count
distributions, dashed curves -- fits obtained using Eqs.~\ref{psf} and
\ref{psf_par}.}
\label{ecf_fits}
\end{figure}

To effectively compute the ${\cal P}(r\!\!\mid\!\!k)$ we first construct a
model PSF. Then, the relevant probability distributions are obtained using
the Monte Carlo method. We note that the analytic form of the PSF should
mimic the radial distribution of counts, while some deviations from the
azimuthal symmetry are of lesser importance. Several simple analytic models
have been tested to fit the observed distribution of counts and it was
found that a function of the form

\begin{equation}
\label{psf}
f(<r) = \frac{r^\alpha}{z + r^\alpha + y\cdot r^{\alpha/2}}
\end{equation}
adequately represents the encircled count fraction (ECF), where $\alpha$,
$z$, and $y$ are parameters depending on the source position and energy.
Fits of Eq.~\ref{psf} to the observed count distribution for several
brightest sources have allowed us to find simple relationships between
these parameters and the off-axis angle. The whole procedure looks as
follows. We have noticed that the PSF parameters $\alpha$, $y$, and $\log
z$ are satisfactorily approximated by linear functions of the off-axis
angle $\theta$:

\begin{equation}
\label{psf_par}
\alpha = a_\alpha \cdot \theta + b_\alpha \\
y = a_y \cdot \theta + b_y \\
\log z = a_z \cdot \theta + b_z\,,
\end{equation}
where $a_s$ and $b_s$ ($s = \alpha$, $y$, $z$) are six parameters which
are substituted in Eq.~\ref{psf} and simultaneously fitted to the observed
distribution of counts in a number ($25$ and $31$ in the S and H band,
respectively) of the strongest point-like sources. In Fig.~\ref{ecf_fits} a
sample of the resultant fits to the observed distributions in the S band
is shown. Since not all the fits are of equal quality, the effects of the
${\cal P}(r\!\!\mid\!\!k)$ approximation are carefully examined. As a good
envelope of errors generated by the imperfections of our fitting
procedures we have constructed two model ${\cal P}(r\!\!\mid\!\!k)$
distributions using the ECF functions systematically wider and narrower by
$15$\,\% as compared to the best fit.

\begin{figure}
\resizebox{\hsize}{!}{\includegraphics{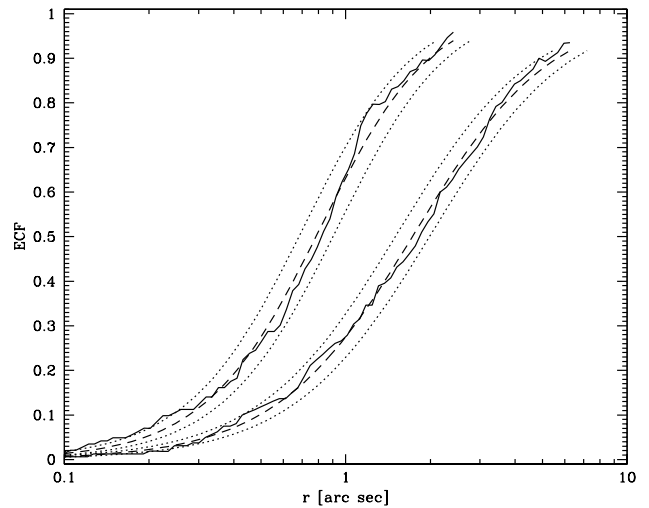}}
\caption{ ECF as in Fig.~\ref{ecf_fits} for sources at $3\farcm 2$ and
$6\farcm 1$ from the field center. Solid curves --  observed count
distributions, dashed curves -- best fits obtained using Eqs.~\ref{psf} and
\ref{psf_par}, dotted curves -- the ECF distributions with the radius $r$
scaled by $\pm 15$\,\%.}
\label{ecf_err}
\end{figure}

Example results of this procedure are illustrated in Fig.\ref{ecf_err},
where the envelope ECFs for two sources at $3\farcm2$ and $6\farcm1$ off
axis angle are shown. Although some deviations are quite large, the
observed ECFs are predominantly contained within the $\pm 15$\,\%
distribution in a wide range of the separations $r$.

In the observational material used in the present investigation the
maximum separations found in the NNST very rarely exceed $2\arcsec$. Thus,
our fits appear adequate for the NNST and it is assumed that the $\pm
15$\,\% limits (indicated by the dotted curves in Fig.\ref{ecf_err})
define the maximum systematic errors associated with the PSF fitting
procedure and will be used to assess uncertainties in our faint source
calculations.

In the Monte Carlo computations of ${\cal P}(r\!\!\mid\!\!k)$ a
population of $10^8$ ``sources'' of $k = 2, 3, ..., k_{\rm max}$ counts
were distributed randomly over the investigated area\footnote{The value
of $k_{\rm max}$ is related the sensitivity threshold and is different
for each energy band; see below.}. The distribution of counts within
each source was randomized according to the model ECF. Then, for each
source a distribution of the nearest neighbor separations was
determined and used to obtain the corresponding amplitudes of ${\cal
P}(r\!\!\mid\!\!k)$. The procedure has been executed for the best fit
and $\pm 15$\,\% ECF distributions.

\subsection{The ``afterglow'' correction}

A charge deposited by a cosmic-ray in the ACIS CCD detector may be
released in two or more time frames generating a sequence of events,
so-called ``afterglow''. The events span typically several seconds and
do not need to occur in consecutive frames \footnote{See
http://cxc.harvard.edu/ciao/why/afterglow.html for details.}.  As a
result, the data contain clumps of counts, which mimic very weak
sources.  Fortunately, the time sequences of such afterglow counts
span short time intervals as compared to exposure times of all the
observations. This allows us to unambiguously identify practically all
the afterglows and remove them from the observation.

\subsection{Pixel randomization}

Positions of counts in the standard ACIS processing are randomized
within the instrument pixel approximated by a square $0\farcs492$ a
side. Since the typical nearest neighbor separations are comparable to
the pixel size, the NNST at small angular scales is smoothed by the
count randomization. The effect is significant and generates most of
the statistical noise in our calculations. To assess uncertainties
introduced by the count randomization, $12$ sets of randomized
observations were produced using the original event data with
non-randomized (integer) counts positions.  Then, the NNST was
determined for each observation and the data were used to obtain the
slope $b$ of the source number counts by means of Eq.~\ref{final}. A
scatter between the $12$ count slope estimates represents the
statistical error of the present method.

\subsection{Strong source removal}

To maximize effect of the weak source population on the NNST, the
strong sources should be removed from the data. The threshold source
flux is defined by $k_{\rm max}$ counts in the Eq.~\ref{final}. Thus,
the value of $k_{\rm max}$ should be set at a level sufficiently large
to ensure that all the sources producing more counts than $k_{\rm
max}$ are found using standard source search criteria. On the other
hand, $k_{\rm max}$ should be small enough to warrant that the number
of sources is adequately represented by the function $N(s)$. 

A catalog of point sources in the AEGIS field is given by
\citet{laird09}.  For each source listed in that paper, a radius
$r_{85}$ enclosing $85$\,\% of counts has been calculated using the
PSF at the source position. Then, several trial values of $k_{\rm
max}$ were applied to assess completeness threshold in the
investigated area. It was found that for $k_{\rm max} = 20$ all the
brighter sources are clearly recognized, while this value is low
enough to allow for statistical treatment of the population of still
weaker sources.

\section{The source counts}

\subsection{The soft band} 

Using the selection criteria given in  Sect.~\ref{observations}, the
area of the field and the total number of counts in the soft band
amount to to $97.5$ sq. arcmin and $73711$, respectively. After the
removal of strong sources according to the procedure described above,
the area is reduced to $92.7$ sq. arcmin and the number of counts to
$48737$.  The average count density amounts to $0.146$ per sq. arcsec.
and the average distance to the nearest neighbor for the random
distribution is equal to$1\farcs22$.  To eliminate a contamination of
the count distribution by strong source photons in the PSF wings, the
removal radius of $\sim 4\cdot r_{85}$ was applied.

The calculations have been performed as follows.  First, the count
distributions were used to calculate the $P(r)$ and $P(r\!\mid\!1)$
probabilities. The distribution of a distances to the nearest neighbor
for each count defines the $P(r)$, while the distribution of the
distances between the randomly distributed points and the nearest
observed count is used to determine $P(r\!\mid\!1)$.  The NNST has
been formulated in the Sect.~\ref{nnst} using the cumulative
probabilities, and estimates of the count slope $b$ obtained from
Eq.~\ref{final} for different separations $r$ are dependent.  To
obtain a set of independent equations, we use the differential
probability distributions $\Delta P(r) = P(r) - P(r+\Delta R)$. In the
calculations the relevant probability distributions have been obtained
over the hole range of the observed separations with $\Delta r =
0\farcs1$. Then, using the Least Square method the best fit value of
the count slope $b$ has been determined for each of the randomized
distributions. Finally, the  average and rms of $b$ was calculated.
The rms obtained in this way results only from the randomization of
counts.

To facilitate comparison of the present results with those available
in the literature, we have adopted from \citet{georgakakis08} the
number counts model for the bright sources. In effect, we fixed
the normalization and the number counts slope in the flux range not
covered by the present analysis, i.e. at fluxes above
$S = 2.9\cdot 10^{-16}$\,cgs or $k_{\rm max} = 20$.
The present calculations provide the slope best estimate below that
flux down to the level determined by the sources generating $2$
counts. The average flux of those sources depends on the number
counts slope and within the power law approximation amounts to 
$S_{\rm w} \approx (2-b)\cdot cf$, where 
$cf = 1.469\cdot 10^{-17}$\,cgs is the average conversion factor
(see Table~2).

\begin{figure}
\resizebox{\hsize}{!}{\includegraphics{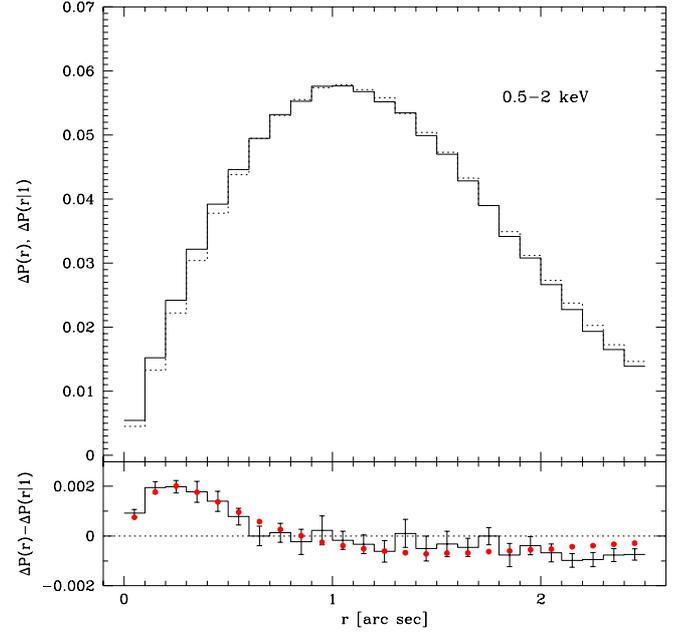}}
\caption{The nearest neighbor probability distributions binned with
$\Delta r = 0\farcs1$ for the observed counts (solid histogram),
and between random and observed counts (dotted histogram);
error bars in the lower panel show $1\sigma$ uncertainties.}
\label{nnpr}
\end{figure}

The  probability distributions $\Delta P(r)$ and $\Delta
P(r\!\mid\!1)$ averaged over $12$ data sets are presented in the upper
panel of Fig.~\ref{nnpr}. A histogram in the lower panel shows the
difference between the both distributions; the error bars indicate
the histogram rms uncertainties obtained from the $12$ pixel
randomization. Dots show the model histogram obtained 
for the NNST best solution slope of $b = 1.595$.

\begin{figure}
\resizebox{\hsize}{!}{\includegraphics{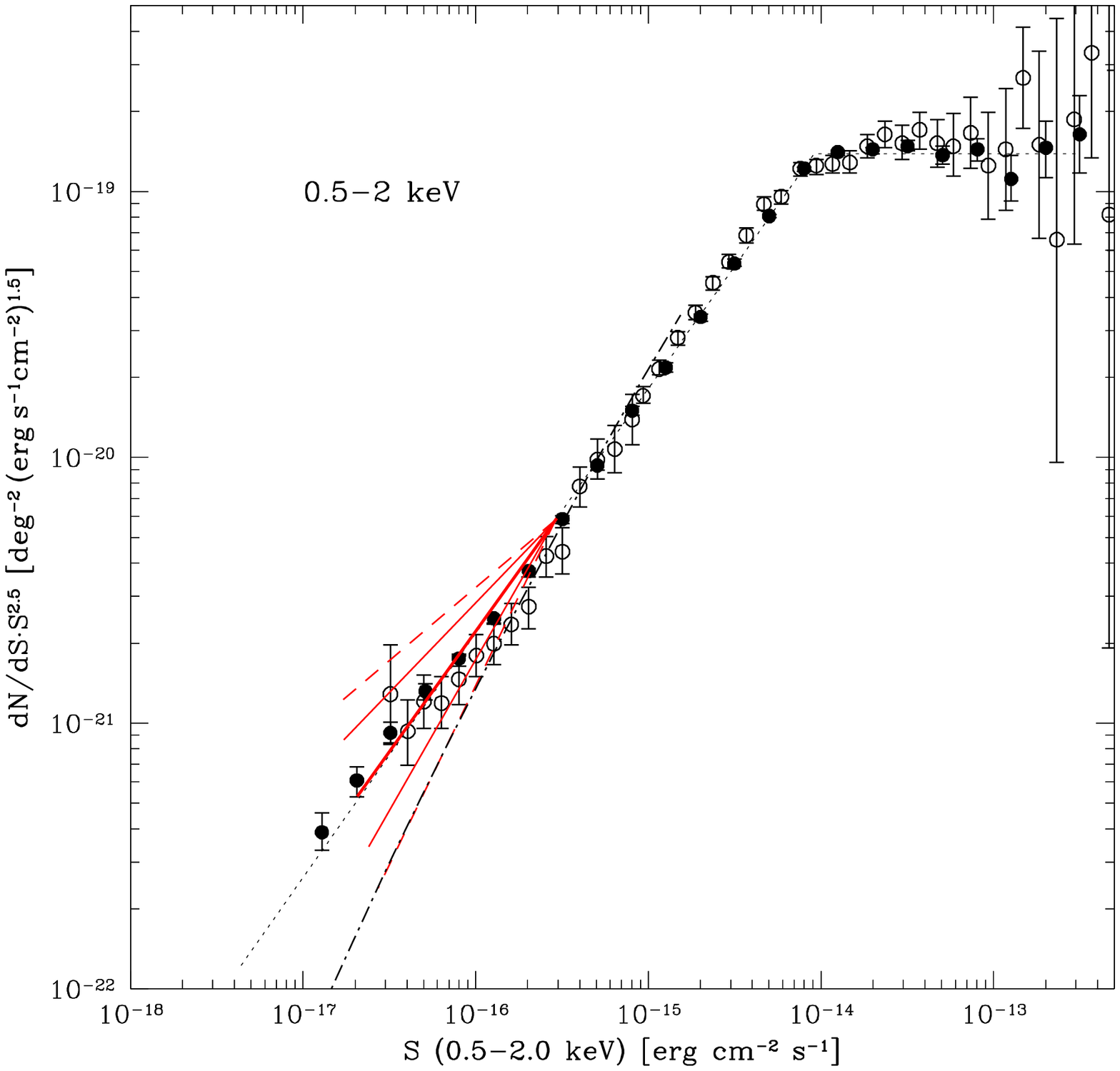}}
\caption{Differential number counts  in the $0.5-2$\,keV band
normalized to the Euclidean slope.
The data points and the dotted line are taken from
\citet{georgakakis08}; dot-dash curve shows the AGN model by
\citet{ueda03}; solid lines - the number counts estimates
based on the NNST: thick solid line - the best power-law fit,
thin solid lines - $1\sigma$ uncertainty range due to the counting
statistics; dashed lines - maximum total uncertainty including
potential errors generated by inaccuracies of the PSF fitting.
The count normalization at the bright end of the NNST model is
fixed at the amplitude given by \citet{georgakakis08}
approximation.}
\label{counts_b1}
\end{figure}

In Fig.~\ref{counts_b1} the results based on the NNST for the S
band are superimposed on the differential number counts presented
by \citet{georgakakis08}. The counts are normalized to the
Euclidean slope of $-2.5$. Full dots and dotted line denote the
counts and model by \citet{georgakakis08}, open circles -- the
counts by \citet{kim07}, and the dot-dash curve -- the predicted
AGN counts from \citet{ueda03}.  The thick line shows the NNST
best fit solution with the slope $b = 1.595$ and the normalization
fixed at $S = 2.9\cdot 10^{-16}$\,cgs. The number counts above
that flux are described by the \citet{georgakakis08} model, i.e.
over a wide range of fluxes the counts are approximated by the
power law with the differential slope of $-1.58$. It is clear that
the NNST solution is in full agreement with the direct source
counts derived from the deep \chandra\ exposures. One should
emphasize that the present NNST solution is based on the
relatively shallow exposure as compared to the
\citet{georgakakis08} data.  In fact, the sources at the low flux
end of the present solution on the average generate in our data
essentially less than $2$ counts each. Evidently, the NNST is
capable to provide a sensitive estimation method for the
population of sources which cannot be recognized as individual
entities. It is worth to note that counts due to the sources
generating the signal, i.e. producing $2\le k \le 20$ photons,
constitute a small fraction of all the counts. In the present
model just $\sim\!1080$ counts or $2.2$\,\% comes from these
sources while the remaining $97.8$\,\% is distributed randomly.

\subsection{Error estimates}

A small fractional contribution of counts produced by sources to
the total number is to be blamed for the rather large statistical
uncertainties of the NNST method. In our case the rms of the slope
$b$ in the S band amounts to $0.230$. The situation is even worse
in the H band, where the rms uncertainty of the slope reaches
$0.34$ (see below).

A question of systematic errors is less straightforward.  Several
potential effects could influence our estimate of the count slope.
The first one, already discussed in Sect.~\ref{psf_sec}, is
associated with our PSF approximation.  The complex shape and
intricate position dependence of the PSF demands approximate
treatment. Our PSF fits inevitably introduce errors.
Unfortunately, an amplitude of these errors is difficult to
assess. Because of that we adopted a cautious approach to this
problem.  Alongside the best fit PSF model we have considered two
ancillary sets of PSFs which delineate the observed count
distribution of the observed strong sources and also confine
possible deviations introduced by our simplified model of the PSF
variations over the field of view. Visual comparison of the actual
count distributions and our model PSF shows that the $\pm 15$\,\%
modification of the PSF width account for any potential
deficiencies of our PSF calculations.

The $\pm 15$\,\% uncertainty of the PSF width introduces a
substantial uncertainty of the slope estimate. For the PSF wider
then the best fit by $15$\,\% we get $b = 1.744 \pm 0.207$, while
for the $15$\,\% narrower, $b = 1.421 \pm 0.267$.  One can
summarize these results as follows. Statistical $1\sigma$ limits
around the best fit solution of $b = 1.595$ are defined as $1.365
< b < 1.825$, while the combined statistical and systematic
uncertainties are $1.154 < b < 1.951$. The statistical
uncertainties are shown in Fig.~\ref{counts_b1} with thin solid
lines and the total uncertainties -- with dashed lines.  One
should note that these ``total'' error estimates are highly
conservative. They have been obtained by simple addition of the
systematic and statistical errors assuming their highest
``possible'' values.

Another source of the slope estimate error is related to the
uncertainty of the number counts normalization. Any modification of
the number of sources at the strong flux end affects the count slope
as well. However, in the relevant flux range the realistic
normalization uncertainties remain small (see Fig.~\ref{counts_b1})
and do not affect considerably the slope uncertainty range.

Variations of the conversion factor over the field of view also do
not play significant role in our slope estimates.  As pointed out in
Sect.~\ref{expmap}, one can incorporate these variations
within the investigated area into uncertainty of
the count normalization $N_o$. Assuming the differential number
counts $N(S) = {\cal N} S^{-b}$
(with flux $S$ in ${\rm erg}\,{\rm cm}^{-2}{\rm s}^{-1}$),
a relationship between the count normalization $N_o$ and
the conversion factor cf is

\begin{equation}
N_o = {\cal N}\cdot {\rm cf}^{\,1-b}\,.
\end{equation}
Thus, for $b=1.595$ and the rms of the conversion factor at the level
of $\sim\!5.9$\,\% (Table~\ref{exposures}), the uncertainty of
$N_o$ amounts to $\sim\!3.5$\,\% and is small in comparison with
the uncertainty of ${\cal N}$ itself.

Substantial variations of the PSF width with the distance from
the telescope axis limit the effectiveness of the NNST method.
This is because the nearest neighbor
distribution $P(r)$ for the Eq.~\ref{final} is estimated using
the actual data, i.e. actual distribution of sources, while the
probability ${\cal P}(r|k)$ is determined by averaging the model
sources over the field of view. This feature introduces
additional uncertainty if the number of sources generating
$k\approx k_{\rm max}$ is small, but should be of lesser
importance in the deep \chandra\ exposures.

\subsection{The hard band}

\begin{figure}
\resizebox{\hsize}{!}{\includegraphics{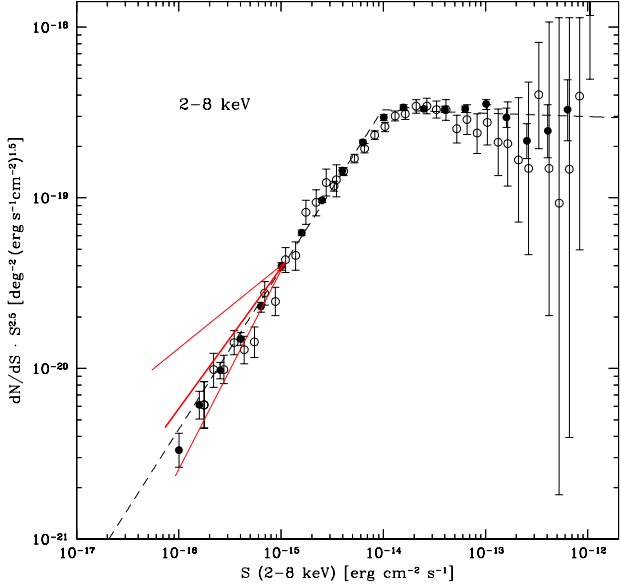}}
\caption{Differential number counts in the $2-8$\,keV band
normalized to the Euclidean slope. The data points and the dashed
line are constructed using the $2-10$\,keV band 
from \citet{georgakakis08}; solid lines - the number
counts estimates based on the NNST: thick solid
line - the best power-law fit, thin solid lines - $1\sigma$
uncertainty range due to the counting statistics.}
\label{counts_b4}
\end{figure}

A fraction of counts produced by discrete sources to all the counts in
the $2-8$\,keV band is substantially smaller than in the $0.5-2$\,keV
band. In effect, the NNST method is less effective in the H than in
the S band, i.e. the slope estimates in the H band are subject to
higher statistical uncertainties. Here we briefly summarize the
results in the H band. 

As a reference data we used those published by \citet{georgakakis08}.
The counts and the analytic fit in the band $2-10$\,keV given in that
paper were converted to the band $2-8$\,keV assuming a power spectrum
with the photon index $\Gamma = -1.4$. 

After the removal of strong sources, the area and the number of counts
used in the NNST amount to $94.4$ sq. arcmin and $101596$,
respectively. Applying the same procedure as in the S band, the best
fit slope  $b = 1.676 \pm 0.340$ was found using the $12$ randomized
distributions. This is shown in Fig.~\ref{counts_b4} with a thick line
and two thin lines. The sources represented by the NNST solution, i.e.
sources producing $2\le k \le 20$ counts, contribute just
$1214^{+979}_{-427}$ photons or $1.19^{+0.96}_{-0.42}$\,\% of all the
counts.  Due to large statistical uncertainty in this band the
NNST does not provide tight limits on the number counts and we
have not plotted here lines representing the range of systematic
uncertainties.

\section{Conclusions}

We have estimated the source number counts in the S band down to
$\sim\!2\cdot 10^{-17}$\,cgs using the merged data with the integrated
exposure time of $\sim\!465$\,ks. This flux level is below the standard
detection threshold for individual sources in the deepest \chandra\
exposures of $2$\,Ms \citep{kim07}.

Our slope estimate below $S = 10^{-16}$\,cgs fits perfectly the actual
discrete source counts determined using such deep observations.
It shows the NNST potential as an effective tool in the investigation
of the extremely weak source population. In the second paper of
this series we plan to apply the NNST method to the \chandra\ Deep
Fields. With a $2$\,Ms exposure the NNST will allow to assess number
counts down to $\sim\!4\cdot 10^{-18}$\,cgs in the $0.5-2$\,keV band
and to $\sim\!2\cdot 10^{-17}$\,cgs in the $2-8$\,keV band, although in
the latter case the expected accuracy of our estimate might be quite
low.

The constraints obtained for the H band are not restrictive. This is
because the contribution of the non X-ray counts increases with energy
and the data become strongly contaminated by the particle background
which effectively ``dilutes'' the counts concentrations produced by
the sources.  In the next paper some prospects to improve the S/N
ratio above $2$\,keV will be explored.

\begin{acknowledgements}
I thank all the people generating the Chandra Interactive Analysis of
Observations software for making a really user-friendly environment.
This work has been partially supported by the Polish KBN grant
1~P03D~003~27.
\end{acknowledgements}

\appendix

\section{Impact of the exposure fluctuations on the NNST}

Effects of the small variations of the exposure over the field of view
on the probability distributions $P(r)$ and $P(r|1)$ in Eq.\ref{final}
are investigated.

The observed distribution of counts is considered here as the
realization of a Poisson process. According to this premise, the
probability of getting an event in the $x-y$ plane of the field of
view (fov) is described by a smooth function $\rho(x,y)$. The
$\rho(x,y)$ amplitude is proportional to the exposure map
(Sect.~\ref{expmap}).  One can define the ``true'' probability density
$\rho_\circ(x,y)$ which would describe the distribution of counts
expected for the perfect instrument with the flat exposure map and -
consequently - a constant conversion factor. Thus, the $\rho$
distribution is related to the $\rho_\circ$  and the fluctuations of
the exposure map, EM:

\begin{equation}
\label{em}
\rho(x,y) = \rho_\circ(x,y)\cdot \frac{{\rm EM}(x,y)}{{\rm EM}_\circ}\,,
\end{equation}
where ${\rm EM}_\circ$ is the exposure map of the perfect instrument.
A natural normalization of the exposure map is assumed:
$\langle{\rm EM}(x,y)\rangle = {\rm EM}_\circ$, where the brackets
$\langle ...\rangle$ denote the averaging over the fov.
Since the distributions $\rho_\circ$ and EM are independent,
$\langle\rho\rangle = \langle\rho_\circ\rangle$.
To quantify the amplitude of the EM fluctuations with a single
parameter $\varepsilon$, we define a function $f(x,y)$:
\begin{equation}
\label{kappa}
\varkappa(x,y) = \frac{{\rm EM}(x,y)}{{\rm EM}_\circ} =
               1 + \varepsilon\,f(x,y)\,,
\end{equation}
in such a way that $\langle f \rangle = 0$ and
$\sigma_{\!\!f}^{\:2} = \langle f^2 \rangle = 1$. We also have
$\langle\varkappa\rangle = 1$ and
$\sigma_{\!\varkappa} = \varepsilon$, where $\sigma_{\!\varkappa}$ is the
rms of $\varkappa$. Since $\sigma_{\!\varkappa} =
\sigma_{\rm EM}/{\rm EM}_\circ$, the data in the
Table~\ref{exposures} show the $\varepsilon$ parameter characterizing
the present observational material remains small; in particular,
it is equal to $0.059$ and $0.062$ for energy bands S and H, respectively.

The expected nearest neighbor probability distributions $P(r|1)$ 
and $P(r)$ are related to the count distribution $\rho$
in a following way:

\begin{equation}
\label{prob1}
P(r|1) = \langle e^{-\pi r^2 \rho(x,y)}\rangle\,,
\end{equation}

\begin{equation}
\label{prob}
P(r) = \frac{1}{\langle \rho \rangle}\;
         \langle\rho(x,y)\cdot e^{-\pi r^2 \rho(x,y)}\rangle\,.
\end{equation}
Substituting Eqs.~\ref{em} and \ref{kappa} into Eqs.~\ref{prob1} and
\ref{prob} one can
expand the distributions $P(r|1)$ and $P(r)$ in powers of the
parameter $\varepsilon$. Since the distributions
of $f(x,y)$ and $\rho_\circ(x,y)$ are uncorrelated,
one finds that the linear term in $\varepsilon$
vanishes.

\end{document}